\documentclass[aps,pra,showpacs,twocolumn,notitlepage,superscriptaddress,nofootinbib,longbibliography]{revtex4-2}

\usepackage{comment}
\usepackage{mathtools}
\mathtoolsset{showonlyrefs}

\usepackage{braket}
\usepackage{amssymb,amsfonts,amsmath,amsthm}
\usepackage[hidelinks]{hyperref}
\usepackage{booktabs}
\usepackage{multirow}
\usepackage[normalem]{ulem}
\usepackage{upgreek}
\usepackage{enumitem}
\usepackage{graphicx}

\def\vecx{\vec{x}}
\def\vecy{\vec{y}}
\def\vecxp{\vec{x}^{\,\prime}}

\def\vecg{\vec{\gamma}}
\def\vecd{\vec{d}}
\def\vecb{\vec{b}}

\def\vecu{\vec{u}}
\def\vecz{\vec{z}}

\def\vecv{\vec{v}}

\newcommand{\gstate}{\gamma}

\def\B{\mathcal{B}}

\def\E{\mathcal{E}}

\def\L{\mathcal{L}}

\def\Rcal{\mathcal{R}}

\def\argmin{\operatorname*{arg\,min}}

\newcommand{\Tr}{\operatorname{Tr}}

\newcommand{\RR}{\mathbb{R}}

\newcommand{\Dn}{\Delta_n}
\newcommand{\Dno}{\Delta_n^{\circ}}

\newcommand{\rhop}{\rho^{\,\prime}}

\newcommand{\fx}{f_{\vecx}}
\newcommand{\gxb}{g_{\vecx,\vecb}}

\newcommand{\Lgx}{\L_{\vecg}(\vecx)}
\newcommand{\Uxg}{L_{\vecx\mid\vecg}}
\newcommand{\Uyg}{L_{\vecy\mid\vecg}}
\newcommand{\Uxpg}{L_{\vecxp\mid\vecg}}
\newcommand{\Uzg}{L_{\vecz\mid\vecg}}

\newcommand{\epsstar}{\varepsilon^{\star}}
\newcommand{\Dtr}{D_{\mathrm{tr}}}
\newcommand{\xstar}{\vecx^{\star}_{\vecd,\vecb}}
\newcommand{\xstary}{\vecx^{\star}_{\vecd,-\vecy}}
\newcommand{\norm}[1]{\left\lVert #1\right\rVert}
\def\majg{\succeq_{\vecg}}

\def\majdb{\succeq_{\vecd,\vecb}}
\def\TO{\underset{\mathbf{TO}}{\longrightarrow}}

\newcommand{\DTO}{\Delta_{\mathrm{TO}}}
\newcommand{\LTOr}{\L_{\mathrm{TO}}(\rho)}

\newcommand{\egap}{\delta_{\epsilon}}
\newcommand{\Edphi}{E_{\vecd,\phi}}

\newtheorem{theorem}{Theorem}
\newtheorem{lemma}{Lemma}
\newtheorem{corollary}{Corollary}

\usepackage{algorithm}
\usepackage{algpseudocode}
\usepackage{float}

\newcommand{\capacityroof}{%
  \mathcal{R}_{\vecx}^{(\vecd,\vecb)}}
\newcommand{\splus}{S^{+}}
\newcommand{\sminus}{S^{-}}
\newcommand{\CapRoof}{\textsc{Capacity-Roof}}
\newcommand{\MaxMin}{\textsc{Maximal-Minimizer}}
\newcommand{\BigO}{O}

\begin{document}

\title{Population Capacity and the Geometry of Thermomajorization}

\begin{abstract}
We characterize finite-dimensional quasiclassical state conversion under thermal operations through population capacities: the maximum population that thermal operations can concentrate on each subset of energy levels. 
We show that these capacities form a normalized, nondecreasing submodular rank function whose base polytope coincides with the thermally reachable population region.
This representation provides complete accessibility constraints and a greedy construction of states maximizing the expectation value of any system observable.
For a prescribed target, exact conversion can be certified by the optimized responses of at most $n-1$ target-adapted binary work-extraction probes.
For approximate conversion, we exploit the Lorenz-curve structure of population capacities to construct, in $O(n^2\log n)$ arithmetic and comparison operations, a reachable state that simultaneously minimizes every convex separable residual error at fixed positive comparison weights.
This state also realizes a common first contact with the corresponding target-centered smoothing regions.
Finally, the optimal trace-distance error equals the largest capacity advantage of the target over the initial state, attained at a target thermomajorization prefix and determined by the same optimized binary-probe responses.
Thus population capacities provide a unified, constructive framework for thermal accessibility, observable optimization, conversion certification, and universal approximation.
\end{abstract}

\author{F. Meroi}
\affiliation{Instituto de Investigaciones Matemáticas Luis A. Santaló, Universidad de Buenos Aires, CONICET, Buenos Aires, Argentina}

\author{C. Massri}
\affiliation{Instituto de Investigaciones Matemáticas Luis A. Santaló, Universidad de Buenos Aires, CONICET, Buenos Aires, Argentina}

\author{M. Losada}
\affiliation{CONICET-Universidad de Buenos Aires, Instituto de Ciencias de la Computaci\'on (ICC), Buenos Aires, Argentina}

\author{G. Bellomo}
\affiliation{CONICET-Universidad de Buenos Aires, Instituto de Ciencias de la Computaci\'on (ICC), Buenos Aires, Argentina}

\author{G. M. Bosyk}
\email{gbosyk@icc.fcen.uba.ar (corresponding author)}
\affiliation{CONICET-Universidad de Buenos Aires, Instituto de Ciencias de la Computaci\'on (ICC), Buenos Aires, Argentina}

\date{\today}

\maketitle


\section{Introduction}

Resource theories provide an operational framework for physical settings in which only a restricted class of states and transformations is freely available \cite{Chitambar2019,Gour2025}.
In quantum thermodynamics, this viewpoint leads to the resource theory of athermality \cite{Lostaglio2019}: for a system with Hamiltonian $H$ coupled to a heat bath at inverse temperature $\beta=1/(k_B T)$, where $k_B$ is the Boltzmann constant, the Gibbs state
\begin{equation}
\gstate_H=\frac{e^{-\beta H}}{\Tr[e^{-\beta H}]}
\label{eq:gibbs-state}
\end{equation}
is free, whereas states out of thermal equilibrium with the bath constitute resources \cite{Janzing2000,Brandao2013,Horodecki2013,Lostaglio2019}.
A paradigmatic class of free transformations is given by thermal operations,
$\Phi(\rho)=\Tr_B[U(\rho\otimes\gamma_{H_B})U^\dagger]$,
where $\gamma_{H_B}$ is the Gibbs state of a bath with arbitrary Hamiltonian $H_B$ at the same inverse temperature $\beta$, and $[U,H+H_B]=0$.
These operations encode access to a thermal environment together with microscopic energy conservation.
Throughout, we work with the closure of thermal operations, including transformations that can be approximated arbitrarily well by thermal operations.

We consider the finite-dimensional quasiclassical sector, with a nondegenerate system Hamiltonian $H=\sum_{i=1}^n E_i\ket{E_i}\bra{E_i}$ and states $\rho=\sum_i x_i\ket{E_i}\bra{E_i}$ and $\sigma=\sum_i y_i\ket{E_i}\bra{E_i}$.
Their population vectors $\vecx$ and $\vecy$ belong to the probability simplex $\Delta_n=\{\vec{p}\in\mathbb{R}^n_{\geq 0}:\sum_i p_i=1\}$.
The Gibbs state \eqref{eq:gibbs-state} is represented by the full-support population vector $\vecg\in\Dno$, with $\gamma_i=e^{-\beta E_i}/Z$ and $Z=\sum_i e^{-\beta E_i}$.
Deterministic convertibility is completely characterized by thermomajorization,
\begin{equation}
\rho\TO\sigma \quad\iff\quad \vecx\majg\vecy .
\label{eq:TO_thermomajorization_intro}
\end{equation}
Here, $\vecx\majg\vecy$ means that the Lorenz curve of $\vecx$ relative to $\vecg$ never lies below that of $\vecy$, providing an exact single-shot criterion finer than the free-energy condition.

While thermomajorization completely decides whether a prescribed conversion is possible, many operational questions concern the entire reachable region: which observable values can be attained, how can conversion be certified, and how closely can an inaccessible target be approximated?
We address these questions through the geometry of the thermally reachable population region
\begin{equation}
\Lgx=\{\vecxp\in\Dn:\vecx\majg\vecxp\},
\label{eq:thermal_lower_set_intro}
\end{equation}
which we call the \textit{population lower set}.
Previous works have investigated its geometry, including thermal cones, extreme points, and degeneracies \cite{Mazurek2019,Junior2022,vomEnde2024}.
Here we identify a submodular structure of the population lower set that enables a unified and constructive treatment of observable optimization, conversion certification, and universal approximation.

Our starting point is the maximum population attainable on a prescribed subset $S$ of energy levels.
For $\vecz\in\RR^n$ and $S\subseteq N=\{1,\ldots,n\}$, write $z(S)=\sum_{i\in S}z_i$.
In particular, $x(S)=\Tr(\rho\Pi_S)$ and $\gamma(S)=\Tr(\gstate_H\Pi_S)$ are the populations assigned to $S$ by the initial and Gibbs states.
We define the \textit{population capacity}
\begin{equation}
C_\rho(S)=\max_{\rhop\in\LTOr}\Tr(\rhop\Pi_S),
\label{eq:pop_cap_def}
\end{equation}
where $\Pi_S=\sum_{i\in S}\ket{E_i}\bra{E_i}$ and $\LTOr$ is the set of states reachable from $\rho$ under the closure of thermal operations.
Since the initial state is quasiclassical, all these states are quasiclassical, and their population vectors are exactly those in $\Lgx$.
All capacities refer to the fixed system Hamiltonian $H$ and inverse temperature $\beta$.
We show that these capacities define a polymatroid rank function, that is, a normalized, nondecreasing, and submodular set function, whose base polytope is exactly $\Lgx$.
Consequently, the capacity inequalities $x'(S)\leq C_\rho(S)$ are not merely necessary but jointly sufficient for thermal accessibility.

The polymatroid structure also provides a constructive solution to observable optimization.
Along any nested chain of subsets of energy levels, the corresponding capacities can be saturated simultaneously by a single reachable state.
This yields a greedy construction of states maximizing the expectation value of any system observable, including the Hamiltonian itself.
Moreover, binary probe Hamiltonians isolate individual capacities within the work-extraction framework of Ref.~\cite{HsiehGessner2026}.
For a prescribed target, the optimized responses of at most $n-1$ target-adapted binary probes suffice to certify exact conversion, establishing the thermomajorization recovery conjectured in that framework for the quasiclassical setting considered here.

When exact conversion is impossible, the same capacity constraints govern how closely the target can be approximated.
We consider residual errors of the form $\sum_i d_i\phi((x'_i-y_i)/d_i)$, where $\vecd\in\Dno$ specifies positive comparison weights and $\phi$ is convex.
General results on least elements of submodular bases guarantee a common optimizer for all such errors at fixed $\vecd$ \cite{Fujishige2005,SchootUiterkamp2025}.
Exploiting the specific Lorenz-curve structure of population capacities, we obtain an explicit construction of this optimizer through a weighted principal sequence of active constraints in $O(n^2\log n)$ arithmetic and comparison operations.
Different choices of comparison weights recover ordinary and Gibbs-weighted errors as well as target-relative $f$-divergences.
For each fixed choice of weights, the distinguished state also realizes a common first contact with the corresponding target-centered smoothing regions \cite{Renes2016,VanDerMeer2017,Chubb2018}.

Finally, we show that the optimal trace-distance conversion error equals the largest excess of target population demand over available capacity, $\max_{S\subseteq N}[y(S)-C_\rho(S)]$.
This deficit is attained at a target thermomajorization prefix and equals the largest positive optimized energy-storage deficit among the same binary probes, divided by their energy gap.
Thus, the capacity constraints that certify exact conversion also quantify its optimal approximation error.

Proofs of the results and the constructive algorithm, together with its complexity analysis, are provided in the Appendices.

\section{Population capacity and polymatroid geometry}

For the fixed Hamiltonian $H$ and inverse temperature $\beta$, every reachable
state is quasiclassical with population vector in $\Lgx$, so the population
capacity~\eqref{eq:pop_cap_def} becomes
\begin{equation}
C_\rho(S)=\max_{\vecxp\in\Lgx}x'(S).
\label{eq:thermal-population-capacity}
\end{equation}

To evaluate these capacities, let $\pi^x$ be a $\vecg$-ordering of $\vecx$, satisfying $x_{\pi^x(1)}/\gamma_{\pi^x(1)} \geq\cdots\geq x_{\pi^x(n)}/\gamma_{\pi^x(n)}$.
These ratios measure occupation relative to equilibrium. 
Define the prefix sets $S_k^{\pi^x}=\{\pi^x(1),\ldots,\pi^x(k)\}$, with $S_0^{\pi^x}=\emptyset$.
The relative Lorenz curve $\Uxg:[0,1]\to[0,1]$ linearly interpolates the nodes $\{(\gamma(S_k^{\pi^x}),x(S_k^{\pi^x}))\}_{k=0}^n$.
It is nondecreasing and concave, and for every $S\subseteq N$,
\begin{equation}
x(S)\leq\Uxg\bigl(\gamma(S)\bigr),
\label{eq:subset-below-curve}
\end{equation}
with equality at the prefixes $S_k^{\pi^x}$.

We now characterize the population capacities and the geometry of the entire thermally reachable region.

\begin{theorem}[Population capacities and polymatroid geometry]
\label{thm:lower-base}
Let $\rho$ be a quasiclassical state with population vector $\vecx\in\Dn$,
and let $\vecg\in\Dno$ be the Gibbs population vector. Then the population
capacity satisfies
\begin{equation}
C_\rho(S)=\Uxg\bigl(\gamma(S)\bigr), \quad \forall S\subseteq N.
\label{eq:subset-support}
\end{equation}
Moreover, $C_\rho$ is a polymatroid rank function with $C_\rho(N)=1$, and
\begin{equation}
\Lgx=B(C_\rho),
\label{eq:lower-equals-base}
\end{equation}
where
\begin{equation}
B(C_\rho)=\left\{\vecz\in\RR^n: z(S)\le C_\rho(S)\ \forall S\subseteq N, \ z(N)=1\right\}.
\end{equation}
\end{theorem}

Theorem~\ref{thm:lower-base} identifies thermal accessibility with a complete system of subset population constraints.
Although each capacity is defined by a separate optimization, these inequalities are jointly sufficient to characterize the reachable region.
Submodularity expresses diminishing returns: enlarging a subset cannot increase the capacity gain from adding a fixed extra energy level.
Moreover, the polymatroid structure guarantees that the capacities along any nested chain of subsets can be saturated simultaneously by a single reachable state.
In particular, the initial state saturates its own thermomajorization prefixes, whereas saturating other subsets may require thermal redistribution.

This simultaneous saturation property provides a constructive solution to observable optimization.
Since every reachable state is diagonal in the energy eigenbasis, only the diagonal entries of an observable contribute to its expectation value.
Ordering these entries selects a nested chain whose capacity bounds are simultaneously attainable.

\begin{corollary}[Optimized observables]
\label{cor:diagonal-observable-support}
Let $A$ be an observable, with diagonal entries $a_i=\bra{E_i}A\ket{E_i}$.
Choose an ordering $\pi^a$ such that $a_{\pi^a(1)}\leq\cdots\leq a_{\pi^a(n)}$, and define $S_k^a=\{\pi^a(k+1),\ldots,\pi^a(n)\}$ for $k=1,\ldots,n-1$, with $S_0^a=N$ and $S_n^a=\emptyset$.

The maximum expectation of $A$ over the thermally reachable set is
\begin{equation}
\max_{\rhop\in\LTOr}\Tr(A\rhop) =a_{\pi^a(1)} +\sum_{k=1}^{n-1} [a_{\pi^a(k+1)}-a_{\pi^a(k)}]C_\rho(S_k^a).
\label{eq:diagonal-observable-support}
\end{equation}
An optimizing population vector is
\begin{equation}
z_{\pi^a(k)}^a =C_\rho(S_{k-1}^a)-C_\rho(S_k^a),
\qquad k=1,\ldots,n.
\label{eq:observable-greedy-state}
\end{equation}
It simultaneously saturates the entire chain:
$z^a(S_k^a)=C_\rho(S_k^a)$ for $k=0,\ldots,n$.
\end{corollary}

The corollary gives both the optimal observable value and a state attaining it.
Once the relative Lorenz curve is available, the construction requires only an ordering of the observable diagonal entries and evaluations of the corresponding prefix capacities, without enumerating
the vertices of the reachable polytope.

As an application, consider the system Hamiltonian $H$, with ordered energies as $E_1<\cdots<E_n$. 
The largest attainable mean system energy is 
\begin{equation}
\begin{aligned}
E_{\max}^{\mathrm{TO}}(\rho;H) &=\max_{\rhop\in\LTOr}\Tr(H\rhop)\\
&=E_1+\sum_{k=1}^{n-1} (E_{k+1}-E_k)C_\rho(S_k^\uparrow),
\end{aligned}
\label{eq:max-physical-energy}
\end{equation}
where $S_k^\uparrow=\{k+1,\ldots,n\}$.
Each spectral gap weights the maximum population attainable above the corresponding energy threshold, and a single reachable state saturates all these capacities simultaneously.

This quantity should not be confused with extractable work: the mean system energy can increase through energy exchange with the bath, consistently with the nonincrease of the nonequilibrium free energy
$F_\beta(\rho)=\Tr(H\rho)-\beta^{-1}S(\rho)$, where $S(\rho)=-\Tr(\rho\log\rho)$.

More generally, choosing binary probe Hamiltonians built from the projectors $\Pi_S$ isolates individual population capacities.
This leads to the work-extraction characterization of exact thermal conversion developed next.

\section{Exact thermomajorization and binary work-extraction probes }

We connect population capacities with the work-extraction framework of Ref.~\cite{HsiehGessner2026}. 
For a state $\tau$ and a probe Hamiltonian $K$, consider the energy-storage enhancement
\begin{equation}
\Delta(\tau;K)=W(\tau;K)-W(\tau;0),
\label{eq:energy-storage-def}
\end{equation}
where $W(\tau;K)=\beta^{-1}S(\tau\,||\,\gstate_K)$ is the asymptotic extractable-work expression, $\gstate_K=e^{-\beta K}/Z_K$, and $Z_K=\Tr(e^{-\beta K})$.
Here $S(\tau\,||\,\eta)=\Tr[\tau(\log\tau-\log\eta)]$ is the quantum relative entropy.
The entropy contributions in \eqref{eq:energy-storage-def} cancel, yielding
\begin{equation}
\Delta(\tau;K) = \Tr(K\tau)+\frac1\beta\log\frac{Z_K}{n}.
\label{eq:HG-delta-linear}
\end{equation}

For a quasiclassical state $\tau$, its optimized energy-storage enhancement under thermal operations is
\begin{equation}
\DTO(\tau;K) =\max_{\tau'\in\L_{\mathrm{TO}}(\tau)} \Delta(\tau';K).
\label{eq:HG-assisted}
\end{equation}
The optimization is over states reachable under thermal operations defined by the fixed system Hamiltonian $H$, whereas $K$ specifies the work-extraction task.

For fixed energy scales $0\leq\epsilon_1<\epsilon_2$, consider the binary probe Hamiltonians
\begin{equation}
K_S=\epsilon_1 I+\egap\Pi_S, \quad \egap=\epsilon_2-\epsilon_1>0.
\label{eq:binary-subset-probe}
\end{equation}
For $\emptyset\subsetneq S\subsetneq N$, each probe assigns energy $\epsilon_2$ to the levels in $S$ and $\epsilon_1$ to those in its complement.
Using Eqs.~\eqref{eq:HG-delta-linear} and \eqref{eq:HG-assisted}, together with the definition of population capacity, we obtain
\begin{equation}
\DTO(\tau;K_S) =\epsilon_1+\egap C_\tau(S) +\frac1\beta\log\frac{Z_{K_S}}{n}.
\label{eq:binary-capacity-response}
\end{equation}
Thus, for a fixed probe, the optimized response depends on the initial state only through its population capacity.
The remaining terms are state independent and cancel when comparing two states.
Moreover, the enhancement is invariant under uniform shifts of the probe Hamiltonian, so only the energy gap $\egap$ matters.

For a prescribed target $\sigma$, let $\pi^y$ be a $\vecg$-ordering of $\vecy$ and define $S_k^{\pi^y}=\{\pi^y(1),\ldots,\pi^y(k)\}$.
Consider the target-adapted binary probes
\begin{equation}
K_k=\epsilon_1 I+\egap\Pi_{S_k^{\pi^y}}, \quad k=1,\ldots,n.
\label{eq:binary-H}
\end{equation}
Since the target saturates its prefix capacities, $C_\sigma(S_k^{\pi^y})=y(S_k^{\pi^y})$, its optimized response is attained by the target itself: $\DTO(\sigma;K_k)=\Delta(\sigma;K_k)$.
The final probe is trivial, $K_n=\epsilon_2 I$, and satisfies $\DTO(\tau;K_n)=0$ for every state $\tau$.

Combining Theorem~\ref{thm:lower-base} with the target-prefix characterization of thermomajorization gives the following complete conversion criterion.
\begin{corollary}[Exact conversion from capacities and binary probes]
\label{cor:exact-conversion}
For quasiclassical states $\rho$ and $\sigma$, the following statements are equivalent:
\begin{enumerate}[label=(\roman*)]
    \item $\vecx\majg\vecy$. \label{item:thermomaj}
    \item $C_\rho(S)\geq y(S)$ for every $S\subseteq N$. \label{item:capacity_charac}
    \item $C_\rho(S)\geq C_\sigma(S)$ for every $S\subseteq N$. \label{item:capacity2_charac}
    \item $C_\rho(S_k^{\pi^y})\geq C_\sigma(S_k^{\pi^y})$ for $k=1,\ldots,n-1$. \label{item:capacity}
    \item $\DTO(\rho;K_k)\geq\DTO(\sigma;K_k)$ for $k=1,\ldots,n-1$.
    \label{item:energy-storage}
\end{enumerate}
\end{corollary}

Exact conversion is therefore characterized either by the initial capacities meeting every target population demand or by the dominance of the entire target capacity profile.
These conditions need only be checked on the target thermomajorization prefixes, where the target saturates its capacities.
Equivalently, at most $n-1$ target-adapted binary probes suffice to certify conversion through their optimized energy-storage responses.
This establishes, in the finite-dimensional quasiclassical setting, the recovery of thermomajorization conjectured in Ref.~\cite{HsiehGessner2026}.

Exact conversion is therefore characterized either by the initial capacities meeting every target population demand or by the dominance of the entire target capacity profile.
Furthermore, it suffices to check the capacity inequalities on the target thermomajorization prefixes, where the target saturates its capacities.
Equivalently, at most $n-1$ target-adapted binary probes suffice to certify conversion through their optimized energy-storage responses.
This establishes, in the finite-dimensional quasiclassical setting, the recovery of thermomajorization conjectured in Ref.~\cite{HsiehGessner2026}.

The same binary probes will also quantify the failure of exact conversion.
After constructing universally optimal reachable approximations, we will show that their largest normalized response deficit equals the minimum trace-distance conversion error.

\section{Universal approximate thermomajorization and first-contact smoothing}

When exact conversion is impossible, we seek the best reachable approximation to the target.
For an error $\E$ between quasiclassical states, the optimal approximate conversion error is
\begin{equation}
\epsstar_{\E}(\rho\TO\sigma) =\inf_{\rhop\in\LTOr}\E(\rhop\,|\,\sigma).
\label{eq:approx-state-conversion}
\end{equation}
Rather than optimizing each error measure separately, we ask whether a single reachable state can simultaneously minimize an entire family of approximation errors.

We consider state errors determined by the population vectors in the fixed energy eigenbasis.
Let $\vecd\in\Dno$ be a vector of positive comparison weights, and let $\phi:I_{\vecd,\vecy}\to\RR$ be a finite convex function with $\phi(0)=0$, where
$I_{\vecd,\vecy} =[-\max_i(y_i/d_i),\,\max_i((1-y_i)/d_i)]$.
Define the convex separable residual error
\begin{equation}
\Edphi(\vecz\,|\,\vecy) =\sum_{i=1}^n d_i  \phi\left(\frac{z_i-y_i}{d_i}\right).
\label{eq:Edphi}
\end{equation}
These errors are nonnegative by Jensen inequality, since $\sum^n_{i=1} d_i=1$ and $\sum^n_{i=1} (z_i-y_i)=0$.
We call an error \textit{faithful} when it vanishes only at $\vecz=\vecy$.

For the corresponding state error $\E_{\vecd,\phi}(\rhop\,|\,\sigma) = \Edphi(\vecxp\,|\,\vecy)$, Eq.~\eqref{eq:approx-state-conversion} reduces to
\begin{equation}
\epsstar_{\vecd,\phi}(\rho\TO\sigma) =\inf_{\vecxp\in\Lgx} \Edphi(\vecxp\,|\,\vecy).
\label{eq:approx-thermo}
\end{equation}
The Gibbs population $\vecg$ determines the reachable region, whereas $\vecd$ determines the comparison geometry and $\phi$ selects a particular error measure within that geometry.

The existence of a common optimizer at fixed comparison weights follows from general results on least elements of submodular bases \cite{SchootUiterkamp2025}.
To formulate this property, it is useful to introduce an arbitrary shift $\vecb\in\RR^n$ and the separable cost
\begin{equation}
F_{\vecd,\vecb,\phi}(\vecz) =\sum_{i=1}^n d_i \phi\left(\frac{z_i+b_i}{d_i}\right),
\label{eq:general-shift-cost}
\end{equation}
where $\phi$ is finite and convex on an interval containing the normalized coordinates under consideration. 
For vectors $\vecu,\vecv \in \RR^n$ with equal total sum, the shifted weighted majorization order \cite{Veinott1971,Joe1990,SchootUiterkamp2025} is
\begin{equation}
\vecu\majdb\vecv \quad\iff\quad F_{\vecd,\vecb,\phi}(\vecu) \geq F_{\vecd,\vecb,\phi}(\vecv)
\quad\forall\phi.
\label{eq:d-majorization}
\end{equation}

Since the reachable region is a submodular base by Theorem~\ref{thm:lower-base}, the theory of least elements gives the following result.

\begin{corollary}[Universal least element]
\label{cor:universal_least}
For every $\vecd\in\Dno$ and $\vecb\in\RR^n$, there is a unique $\xstar\in\Lgx$ such that $\vecxp\majdb\xstar$ for all $\vecxp\in\Lgx$.
Equivalently,
\begin{equation}
\bigcap_{\phi\ \mathrm{convex}}
\argmin_{\vecxp\in\Lgx} F_{\vecd,\vecb,\phi}(\vecxp) =\{\xstar\}.
\label{eq:db-universal-min}
\end{equation}
\end{corollary}

We now exploit the specific structure of population capacities to construct this universal least element.
Define the shifted capacity function
\begin{equation}
\gxb(S)=C_\rho(S)+b(S),
\label{eq:gxb}
\end{equation}
whose base satisfies
\begin{equation}
B(\gxb)=\Lgx+\vecb.
\label{eq:translated-base}
\end{equation}
The modular shift preserves submodularity, although $\gxb$ need not be nondecreasing.

Consider the discrete capacity roof
\begin{equation}
\Rcal_{\vecx}^{(\vecd,\vecb)} = \{(d(S),\gxb(S)):S\subseteq N\}
\label{eq:roof-set}
\end{equation}
and its greatest convex minorant, running from $(0,0)$ to $(1,1+b(N))$.
The weighted principal sequence \cite{Fujishige1980,Fujishige2005} provides nested sets $\emptyset=S_0\subsetneq S_1\subsetneq\cdots\subsetneq S_m=N$ whose roof points are its successive vertices.
For the blocks $B_j=S_j\setminus S_{j-1}$, the strictly increasing slopes are
\begin{equation}
\lambda_j = \frac{\gxb(S_j)-\gxb(S_{j-1})}{d(B_j)}, \quad j=1,\ldots,m.
\label{eq:slopes}
\end{equation}

\begin{theorem}[Constructive formula for the least element]
\label{thm:construction}
Let $\vecd\in\Dno$ and $\vecb\in\RR^n$, and let
$\emptyset=S_0\subsetneq S_1\subsetneq\cdots\subsetneq S_m=N$ be the weighted
principal sequence of $\gxb$, with blocks $B_j=S_j\setminus S_{j-1}$ and slopes
$\lambda_j$ given by Eq.~\eqref{eq:slopes}. Then the universal least element
$\xstar$ of Corollary~\ref{cor:universal_least} is
\begin{equation}
x^\star_{\vecd,\vecb,i}=d_i\lambda_j-b_i, \quad i\in B_j.
\label{eq:construction-main}
\end{equation}
Moreover, it saturates the population capacities of the cumulative sets:
\begin{equation}
x^\star_{\vecd,\vecb}(S_j)=C_\rho(S_j), \quad j=0,\ldots,m.
\label{eq:principal-capacity-saturation}
\end{equation}
\end{theorem}

For population capacities, the Lorenz-curve representation enables an explicit construction of the weighted principal sequence without enumerating the exponentially many roof
points or invoking general-purpose submodular minimization.
Appendix~\ref{app:algorithm} provides an algorithm requiring $O(n^2\log n)$ arithmetic and comparison operations.
For Gibbs weights ($\vecd=\vecg$) with arbitrary shift $\vecb$, or target weights ($\vecd=\vecy\in\Dno$) with $\vecb=-\vecy$, the construction further simplifies to
the lower convex hull of $n+1$ suitably ordered prefix points, reducing the complexity to $O(n\log n)$ (Corollary~\ref{cor:supp-prefix}).

We now specialize the construction to approximation of the target $\sigma$ by setting $\vecb=-\vecy$.
The shifted capacity ~\eqref{eq:gxb} becomes 
\begin{equation}
g_{\vecx,-\vecy}(S)=C_\rho(S)-y(S),
\label{eq:target-capacity-surplus}
\end{equation}
which measures the capacity surplus relative to the target population demand.
A negative value identifies a violated accessibility constraint.
Since $g_{\vecx,-\vecy}(\emptyset) =g_{\vecx,-\vecy}(N)=0$, the corresponding capacity roof has endpoints $(0,0)$ and $(1,0)$.
Within each principal block, $x^\star_{\vecd,-\vecy,i}-y_i=d_i\lambda_j$: the normalized population residual is constant, whereas the cumulative principal sets identify
saturated capacity constraints.

For this choice of shift, $F_{\vecd,-\vecy,\phi}(\vecxp) =\Edphi(\vecxp\,|\,\vecy)$.
Corollary~\ref{cor:universal_least} therefore shows that the same reachable vector $\xstary$ attains the optimal error for every admissible
convex generator:
\begin{equation}
\epsstar_{\vecd,\phi}(\rho\TO\sigma)
=\Edphi(\xstary\,|\,\vecy).
\label{eq:optimal-error}
\end{equation}
The universal optimizer depends on the comparison weights $\vecd$, but not on the convex generator $\phi$ within that geometry.

This result also admits a target-centered smoothing interpretation \cite{Renes2016,VanDerMeer2017,Chubb2018,Horodecki2018,Massri2020}.
For each faithful error, define the convex tolerated region
\begin{equation}
\B_\varepsilon^{\vecd,\phi}(\vecy) =\{\vecz\in\Dn: \Edphi(\vecz\,|\,\vecy)\leq\varepsilon\}.
\label{eq:smoothing-ball}
\end{equation}

\begin{corollary}[Universal first-contact smoothing]
\label{cor:smoothing-equivalence}
Fix $\vecd\in\Dno$. 
For every faithful error $\Edphi$, let $\epsstar=\epsstar_{\vecd,\phi}(\rho\TO\sigma)$.
Then
\begin{equation}
\Lgx\cap\B_\varepsilon^{\vecd,\phi}(\vecy) =\emptyset\qquad(0\le\varepsilon<\epsstar),
\end{equation}
whereas
\begin{equation}
\xstary\in\Lgx\cap \B_{\epsstar}^{\vecd,\phi}(\vecy).
\label{eq:smoothing-equivalence}
\end{equation}
\end{corollary}

Thus, for fixed comparison weights, the same reachable state realizes the first contact for every faithful error in the family.
Strictly convex generators yield a unique contact state, whereas other generators may admit additional minimizers.
Changing the comparison weights may select a different distinguished contact state.

Three choices illustrate the comparison geometries.

\textit{Ordinary residual geometry.}---
Uniform weights, $\vecd=\vecu=(1/n,\ldots,1/n)$, measure absolute population deviations on the same scale. 
For $\phi_p(t)=|t|^p$, $1\leq p<\infty$, we obtain
\begin{equation}
\label{eq:norm_p_uniform}
E_{\vecu,\phi_p}(\vecxp\,|\,\vecy) =n^{p-1}\norm{\vecxp-\vecy}_p^p.
\end{equation}
Thus, $\vecx^\star_{\vecu,-\vecy}$ simultaneously minimizes all $\ell_p$ errors for $1\leq p<\infty$.

\textit{Gibbs-weighted residual geometry.}---
Choosing $\vecd=\vecg$ measures deviations in units of equilibrium occupation:
\begin{equation}
\label{eq:norm_p_gibbs}
E_{\vecg,\phi_p}(\vecxp\,|\,\vecy) =\sum_{i=1}^n \frac{|x'_i-y_i|^p}{\gamma_i^{p-1}}.
\end{equation}
Accordingly, $\vecx^\star_{\vecg,-\vecy}$ simultaneously minimizes these Gibbs-weighted $\ell_p$ errors for all $1\leq p<\infty$.

\textit{Target-relative geometry.}---
For a full-support target $\vecy\in\Dno$, choosing $\vecd=\vecy$ measures deviations relative to the target populations.
A finite convex $f:[0,\infty)\to\RR$ with $f(1)=0$
and $\phi_f(t)=f(1+t)$ gives
\begin{align}
\label{eq:norm_p_target}
E_{\vecy,\phi_f}(\vecxp\,|\,\vecy)
&=\sum_{i=1}^n y_i f(x'_i/y_i)\\
&=D_f(\vecxp\,||\,\vecy).
\end{align}
Thus, $\vecx^\star_{\vecy,-\vecy}$ simultaneously minimizes all Csisz\'ar $f$-divergences with finite convex generators, including relative entropy for $f(s)=s\log s$, with $0\log0=0$.

\begin{figure*}[t]
\centering
\includegraphics[width=\textwidth]{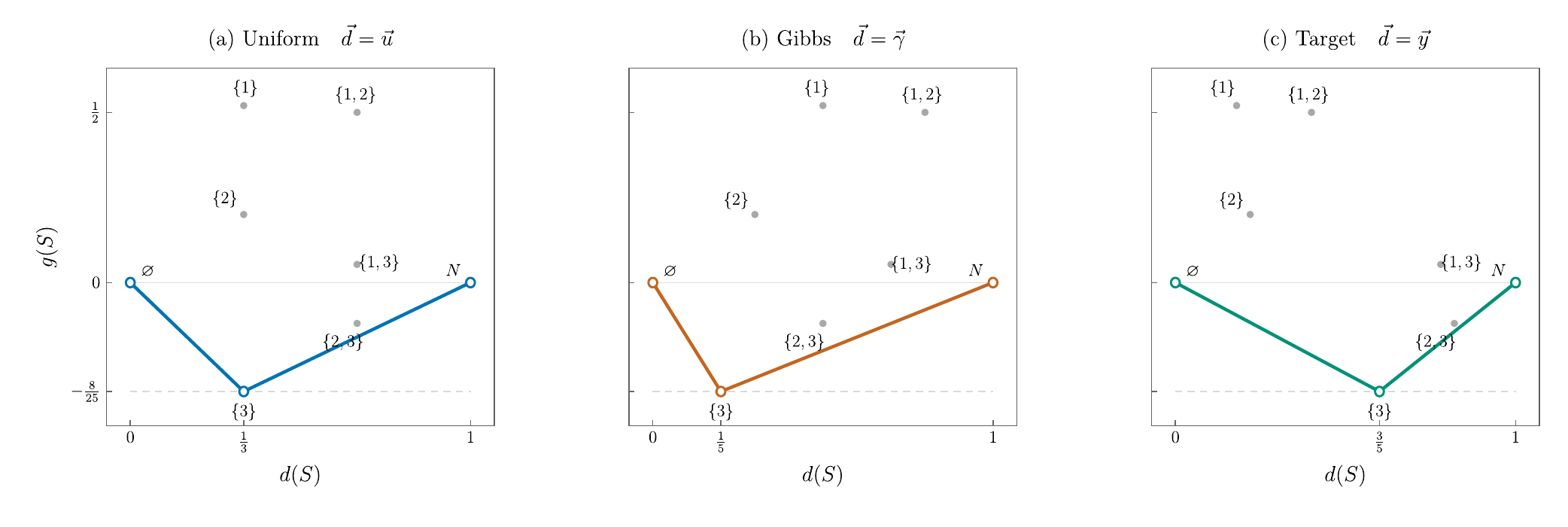}
\caption{Greatest convex minorant (GCM) of the capacity roof for three comparison  geometries from which $\xstar$ is obtained: (a) uniform, $\vec d=\vec u$; (b) Gibbs-weighted,
  $\vec d=\vec\gamma$; and (c) target-weighted, $\vec d=\vec y$.   
  Each panel shows the eight roof points $(d(S),C_\rho(S)-y(S))$ for the same initial, Gibbs, and target population vectors $\vecx=(7/10,1/5,1/10)$, $\vecg=(1/2,3/10,1/5)$, and $\vecy=(9/50,11/50,3/5)$, respectively. 
  Gray dots indicate the remaining subsets; colored open circles mark the principal sets $\emptyset$, $\{3\}$, and $N$,  connected by the greatest convex minorant (solid line).
  Although the principal chain is identical, the position of its  intermediate vertex changes from $d(\{3\})=1/3$ to $1/5$ and $3/5$,
  respectively, leading to different slopes and universal optimizers.}
\label{fig:supp-gcm-geometries}
\end{figure*}

Figure~\ref{fig:supp-gcm-geometries} illustrates the construction of GCM for the three comparison geometries from which $\xstar$ is obtained. 
For the initial, Gibbs, and target population vectors $\vecx=(7/10,1/5,1/10)$, $\vecg=(1/2,3/10,1/5)$, and $\vecy=(9/50,11/50,3/5)$, respectively, the GCM has the same principal chain, $\emptyset\subsetneq\{3\}\subsetneq N$, in all three cases. 
Its intermediate vertex corresponds to the capacity deficit $C_\rho(\{3\})-y(\{3\})=-8/25$.
Although the principal sets coincide, their horizontal positions depend on the comparison weights, leading to different slopes and universal optimizers.

Although the three geometries select different universal optimizers, they share the same optimal trace-distance error. 
This independence of the comparison weights is a general property of the $\ell_1$ error. 
Precisely, for $\phi_1(t)=|t|$, the comparison weights cancel: 
\begin{equation}
\label{eq:norm_1_all}
E_{\vecd,\phi_1}(\vecxp\,|\,\vecy)=\norm{\vecxp-\vecy}_1.   
\end{equation}
Consequently, every distinguished state $\xstary$ is an $\ell_1$-optimal approximant, regardless of the choice of positive comparison weights.
For non-strictly convex errors such as the trace distance, the optimal approximant need not be unique.
Although different comparison weights select distinguished optimal states, their universal optimality does not imply that any of them thermomajorizes all other optimal approximants.

\begin{figure}[t]
\centering
\includegraphics[width=.99\linewidth]{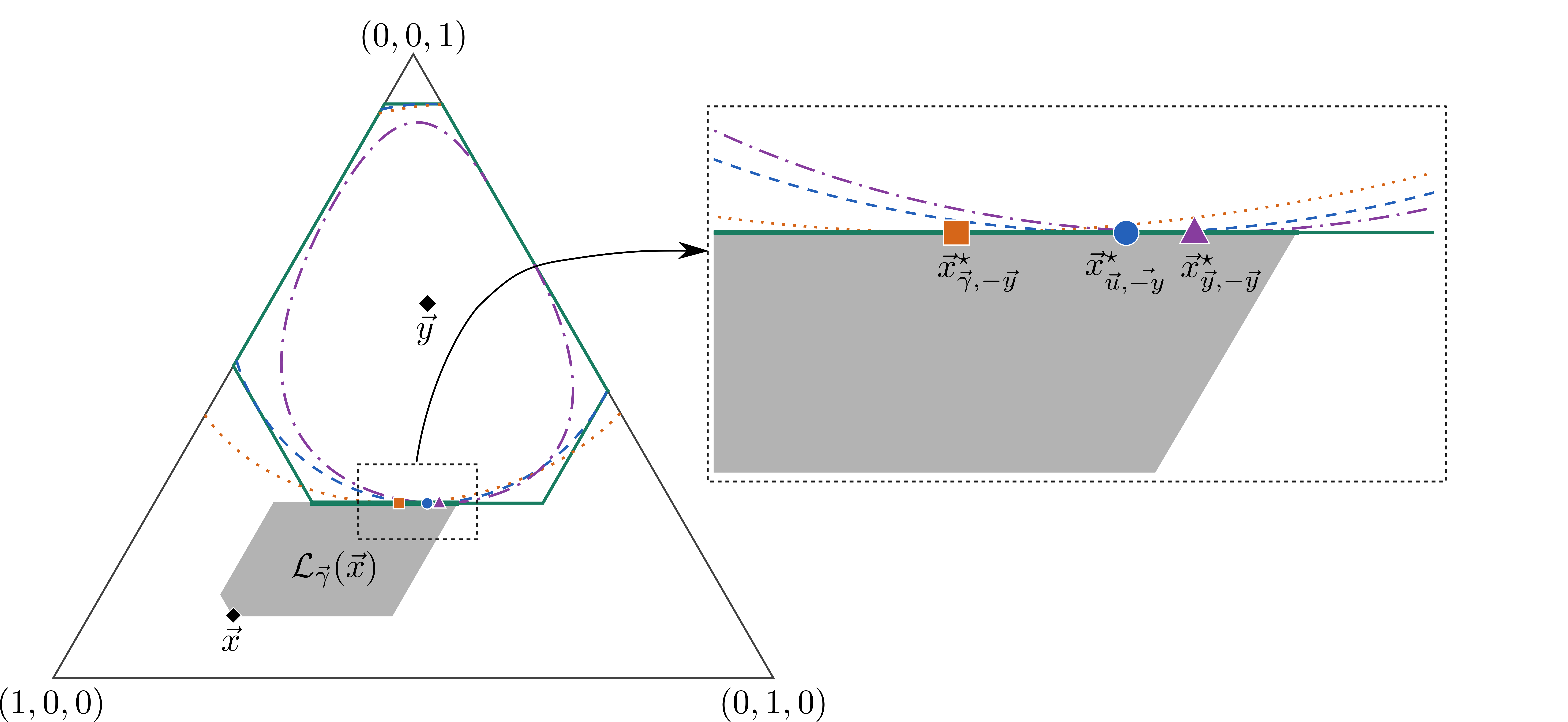}
\caption{\textbf{Universal first-contact geometry.}
Population lower set $\Lgx$ (shaded polygon) for $\vecx=(7/10,1/5,1/10)$, $\vecg=(1/2,3/10,1/5)$, and $\vecy=(9/50,11/50,3/5)$.
Black diamonds mark the initial and target population vectors.
Curves delimit tolerated regions: $\ell_1$ region ($\epsstar_{\vecd,\phi_1}=16/25$, green);
the weighted quadratic regions ($\epsstar_{\vecu,\phi_2}=0.4608$, blue;
$\epsstar_{\vecg,\phi_2}=0.64$, orange); 
and the relative-entropy region ($\epsstar_{\vecy,\phi_f}\simeq0.209807$, purple),
where $\phi_2(t)=t^2$ and $\phi_f(t)=(1+t)\log(1+t)$, with $0\log0=0$.
Inset: the corresponding distinguished states are $\vecx^\star_{\vecu,-\vecy}=(17/50,19/50,7/25)$ (blue circle),
$\vecx^\star_{\vecg,-\vecy}=(19/50,17/50,7/25)$ (orange square), and
$\vecx^\star_{\vecy,-\vecy}=(81/250,99/250,7/25)$ (purple triangle).
Each uniquely minimizes its associated strictly convex error, whereas all three lie on the common $\ell_1$-optimal contact segment.}
\label{fig:thermal-first-contact}
\end{figure}

Figure~\ref{fig:thermal-first-contact} illustrates the different approximants selected by $\vecd=\vecu,\vecg,\vecy$, all belonging to the same $\ell_1$-optimal contact segment.
For this example, $C_\rho(\{3\})=7/25$, whereas $y(\{3\})=3/5$.
Their difference, $8/25$, equals half the optimal $\ell_1$ error shown in the figure.
The next section establishes this relation between trace-distance error and capacity deficits in general.

\section{Trace-distance error from capacity and work-extraction deficits}

For quasiclassical states, the trace distance reduces to one-half the $\ell_1$-distance between their populations: $\Dtr(\rhop,\sigma) =\frac12\Tr|\rhop-\sigma| =\frac12\norm{\vecxp-\vecy}_1$.
For $\phi_1(t)=|t|$, Eq.~\eqref{eq:optimal-error} therefore gives
\begin{equation}
\begin{aligned}
\Dtr^\star(\rho\TO\sigma)
&=\min_{\rhop\in\LTOr}\Dtr(\rhop,\sigma)\\
&=\frac12\epsstar_{\vecd,\phi_1}(\rho\TO\sigma),
\end{aligned}
\label{eq:optimal-trace-error}
\end{equation}
independently of the comparison weights $\vecd$.
Every distinguished approximant $\xstary$ attains this minimum, although the selected state may depend on $\vecd$.

The characterization of approximate conversion in Ref.~\cite[Prop.~6]{Renes2016} identifies this error with the maximal vertical excess of the target relative Lorenz curve over the initial one.
We now express this error through population-capacity deficits and optimized binary-probe responses, and connect its attainment to the universal approximants constructed above.
\begin{theorem}[Capacity and work-extraction characterization of trace-distance error]
\label{thm:trace-distance-work-extraction}
For quasiclassical states $\rho$ and $\sigma$,
\begin{subequations}
\begin{align}
\Dtr^\star(\rho\TO\sigma) &=\max_{S\subseteq N}   [y(S)-C_\rho(S)] \label{eq:trace-capacity-target-demand}\\
&=\max_{S\subseteq N}   [C_\sigma(S)-C_\rho(S)] \label{eq:trace-capacity-deficit}\\
&=\max_{k \in N}   [C_\sigma(S_k^{\pi^y})-C_\rho(S_k^{\pi^y})] \label{eq:trace-target-capacity-gap}\\
&=\frac1{\egap}\max_{k \in N}   [\DTO(\sigma;K_k)-\DTO(\rho;K_k)]. \label{eq:trace-from-work}
\end{align}
\end{subequations}
Here the target-adapted probes $K_k$ are defined in Eq.~\eqref{eq:binary-H} with energy gap $\egap>0$.
\end{theorem}

The theorem shows that the same capacity constraints that determine exact convertibility also quantify its failure.
The optimal error is the largest excess of target population demand over available capacity, equivalently the largest advantage of the target capacity profile.
This deficit is attained at a target thermomajorization prefix and equals the largest normalized deficit in optimized work-extraction responses.
Thus, the error can be determined from the same at most $n-1$ nontrivial binary probes used for exact conversion certification.

The universal approximant provides a constructive certificate of optimality.
For any fixed $\vecd\in\Dno$, let $\vecv=\xstary$ and define $S_-=\{i:v_i<y_i\}$.
By Eq.~\eqref{eq:construction-main}, the normalized residuals satisfy $v_i-y_i=d_i\lambda_j$ on each principal block $B_j$.
Since the slopes are strictly increasing, $S_-$ is the union of the negative-slope blocks and therefore a cumulative principal set.
Theorem~\ref{thm:construction} then gives $v(S_-)=C_\rho(S_-)$, so
\begin{equation}
\Dtr^\star(\rho\TO\sigma) =y(S_-)-C_\rho(S_-).
\label{eq:trace-active-capacity-witness}
\end{equation}
Hence the construction yields both a reachable state attaining the optimal error and a saturated capacity constraint certifying its optimality.

In particular, the optimal trace-distance error vanishes exactly when $\vecx\majg\vecy$, recovering the exact conversion criterion of Corollary~\ref{cor:exact-conversion}.

\section{Discussion}

Population capacities organize quasiclassical thermal conversion through a common reachable geometry.
Their simultaneous saturation along nested sets of energy levels links accessibility constraints to attainable operational bounds.
Observable optimization selects these sets through ordered diagonal entries, exact conversion through target thermomajorization prefixes, and approximation
through a weighted principal sequence of active capacity constraints.

The significance of this representation extends beyond a geometric reformulation of thermomajorization.
While thermomajorization already provides a complete criterion for exact state conversion, the identification
of the reachable region as a polymatroid base makes available a common set of constructive optimization tools.
In particular, the existence of universal least elements follows from general results on submodular bases \cite{Fujishige2005,SchootUiterkamp2025}, whereas the specific Lorenz-curve structure of population capacities enables their explicit construction in $O(n^2\log n)$ operations.
For fixed comparison weights, the resulting state simultaneously minimizes an entire family of convex separable residual costs.
Moreover, the capacity representation connects exact conversion certification with optimal approximation: the same binary work-extraction probes
determine whether conversion is possible and quantify its failure in trace distance.
The universal construction also provides an optimal reachable state together with a saturated capacity constraint certifying the attained error.

The possible nonuniqueness of optimal approximants for non-strictly convex errors raises a further
resource-theoretic question. In particular, does the set of trace-distance-optimal
reachable states contain a greatest element under thermomajorization, namely an optimal state that
thermomajorizes every other optimal approximant?
If not, characterizing its maximal elements and determining whether they can be obtained through suitable comparison weights would provide a finer understanding of approximate thermal conversion.

The same capacity geometry also suggests applications to the optimization of ergotropy \cite{Allahverdyan2004} over thermally reachable states, as encountered in bath-assisted work extraction and thermal machines \cite{Biswas2022}.
For energy-diagonal states, ergotropy is the maximum of finitely many linear population functionals associated with energy-level permutations. 
While the polymatroid greedy construction optimizes each such functional, exploiting this structure to avoid an exhaustive optimization over permutations remains an interesting direction for future work.

Finally, our analysis is restricted to finite-dimensional quasiclassical states under the closure of thermal operations.
Extensions to states with energy coherence or degenerate energy blocks require additional structure beyond the population description developed here.
More broadly, identifying when the reachable regions of other resource theories admit submodular capacity representations could extend the same geometric and constructive approach to new classes of resource transformations.


\begin{acknowledgments}
This work is partially supported by Consejo Nacional de Investigaciones Cient\'ficas y T\'ecnicas CONICET (Argentina) through project PIP 2023-2025 Nro. 0135, and Universidad de Buenos Aires UBA (Argentina) through project UBACyT Programación Científica 2026 and by the European Union through the MSCA SE project QCOMICAL (Grant Agreement ID:101182520).
\end{acknowledgments}

\section*{Author contributions}
All authors contributed to the development of the ideas and results presented in the manuscript. 
GMB supervised the project and prepared the initial draft of the manuscript. 
All authors reviewed and approved the final version of the manuscript and take full responsibility for its content, including the correctness of the mathematical results.
Large language models (LLMs) were used to assist with manuscript drafting, checking mathematical derivations, numerical results and figures.

\appendix

\section{Submodularity and base polytopes}
\label{app:base_polytope}

A set function $f:2^N\to\RR$ is submodular if
\begin{equation}
f(A)+f(B)\geq f(A\cup B)+f(A\cap B),
\qquad A,B\subseteq N.
\label{eq:submodular-def}
\end{equation}
Equivalently, it satisfies the diminishing-returns condition
\begin{equation}
f(A\cup\{i\})-f(A) \geq f(B\cup\{i\})-f(B)
\end{equation}
for $A\subseteq B\subseteq N$ and $i\notin B$.
The function is normalized if $f(\emptyset)=0$
and nondecreasing if $A\subseteq B$ implies $f(A)\leq f(B)$.
A normalized, nondecreasing submodular function is a polymatroid rank function.

For a normalized submodular function $f$, its base polytope is defined by
\cite{Fujishige2005}
\begin{equation}
B(f)=\left\{\vecz\in\RR^n: z(S)\leq f(S)\ \forall S\subseteq N, \quad z(N)=f(N)\right\}.
\label{eq:base-polytope}
\end{equation}
When $f$ is also nondecreasing, $B(f)$ is a polymatroid base polytope and all its vectors have nonnegative components.
For a modular function $b(S)=\sum_{i\in S}b_i$, the shifted function $g=f+b$ remains normalized and submodular, although it need not be nondecreasing.
Its base is the translation $B(g)=B(f)+\vecb$.

We use the greedy property of submodular bases \cite[Prop.~3.2]{Bach2013}.
For any permutation $\pi$, let $P_k=\{\pi(1),\ldots,\pi(k)\}$, with $P_0=\emptyset$.
The greedy vector is defined by
\begin{equation}
w_{\pi(k)}=f(P_k)-f(P_{k-1}), \qquad k=1,\ldots,n.
\end{equation}
It belongs to $B(f)$ and simultaneously saturates the corresponding nested chain: $w(P_k)=f(P_k)$ for all $k$.
Moreover, if $a_{\pi(1)}\geq\cdots\geq a_{\pi(n)}$, this vector maximizes $\sum_i a_i z_i$ over $\vecz\in B(f)$.
In particular, choosing $a_i=\mathbf{1}_S(i)$, where $\mathbf{1}_S$ is the indicator function of $S$, and a permutation whose first $|S|$ elements form $S$, the greedy vector saturates
$w(S)=f(S)$, yielding
\begin{equation}
\max_{\vecz\in B(f)}z(S)=f(S).
\label{eq:em-subset-support}
\end{equation}

\section{Proofs}

\subsection{Population capacity and observable optimization}

\begin{proof}[Proof of Theorem~\ref{thm:lower-base}]
Let us define $\fx=\Uxg \circ \gamma$. First we prove that $\fx$ is a polymatroid rank function. 
The endpoint values and monotonicity of $\Uxg$ give $\fx(\emptyset)=0$, $\fx(N)=1$, and monotonicity of $\fx$.
For $A\subseteq B$ and $i\notin B$, write $a=\gamma(A)$, $b=\gamma(B)$, and $c=\gamma_i$.
Since $0\leq a\leq b$ and $b+c\leq1$, concavity gives
\begin{equation}
\Uxg(a+c)-\Uxg(a)\geq\Uxg(b+c)-\Uxg(b),
\label{eq:concave-increments}
\end{equation}
which proves submodularity by diminishing returns.

We prove now both inclusions in Eq.~\eqref{eq:lower-equals-base}.

If $\vecxp\in\Lgx$, the subset bound and thermomajorization imply
\begin{equation}
x'(S)\leq \Uxpg(\gamma(S))\leq\Uxg(\gamma(S))=\fx(S).
\label{eq:lower-to-base}
\end{equation}
Together with $x'(N)=1=\fx(N)$, this gives $\Lgx\subseteq B(\fx)$.

Conversely, if $\vecz\in B(\fx)$, then
$z_i=1-z(N\setminus\{i\})\geq1-\fx(N\setminus\{i\})\geq0$.
Thus $\vecz\in\Dn$.
At each node of its relative Lorenz curve, with $\pi^z$ a $\vecg$-ordering,
\begin{equation}
 \Uzg(\gamma(S_k^{\pi^z})) =z(S_k^{\pi^z})\leq\fx(S_k^{\pi^z}) =\Uxg(\gamma(S_k^{\pi^z})).
\label{eq:breakpoint-comparison}
\end{equation}
Between consecutive nodes, $\Uzg$ is affine and $\Uxg$ is concave, so this inequality holds throughout $[0,1]$.
Hence $\vecz\in\Lgx$, proving $\Lgx=B(\fx)$.

Finally,  applying the greedy subset-support identity \eqref{eq:em-subset-support} to this base gives
\begin{equation}
C_\rho(S) =\max_{\vecxp\in\Lgx}x'(S) =\fx(S) =\Uxg(\gamma(S)),
\end{equation}
as claimed.

\end{proof}

\begin{proof}[Proof of Corollary~\ref{cor:diagonal-observable-support}]
Since every reachable state is quasiclassical, $\Tr(A\rhop)=\sum_i a_i x'_i$.
For any $\vecz\in\Lgx$, summation by parts in the nondecreasing order $\pi^a$ gives
\begin{equation}
\sum_{i=1}^n a_i z_i=a_{\pi^a(1)} +\sum_{k=1}^{n-1} \bigl[a_{\pi^a(k+1)}-a_{\pi^a(k)}\bigr] z(S_k^a).
\end{equation}
All coefficient differences are nonnegative, and $z(S_k^a)\leq C_\rho(S_k^a)$.
This yields the upper bound in Eq.~\eqref{eq:diagonal-observable-support}.

To show attainability, apply the greedy construction of Appendix~\ref{app:base_polytope} to $C_\rho$ using the reverse ordering
$\pi^a(n),\ldots,\pi^a(1)$.
Its nested prefixes are precisely the sets $S_k^a$, taken in reverse order, and the resulting greedy vector has components
\begin{equation}
z_{\pi^a(k)}^a =C_\rho(S_{k-1}^a)-C_\rho(S_k^a),
\qquad k=1,\ldots,n.
\end{equation}
By the greedy property, $\vecz^a\in B(C_\rho)=\Lgx$ and $z^a(S_k^a)=C_\rho(S_k^a)$ for every $k$.
It therefore attains the upper bound, proving both the optimal value and the optimizing-state construction.
\end{proof}

\subsection{Characterization of exact thermal conversion}

\begin{proof}[Proof of Corollary~\ref{cor:exact-conversion}]
By Theorem~\ref{thm:lower-base}, the population lower set satisfies $\Lgx=B(C_\rho)$.
Since $\vecy\in\Dn$ and $C_\rho(N)=1$, condition \ref{item:capacity_charac} is equivalent to $\vecy\in\Lgx$, proving \ref{item:thermomaj} $\iff$ \ref{item:capacity_charac}.

If \ref{item:thermomaj} holds, then $\Uxg(t)\geq\Uyg(t)$ for all $t\in[0,1]$.
Evaluating at $t=\gamma(S)$ and using Eq.~\eqref{eq:subset-support} gives $C_\rho(S)\geq C_\sigma(S)$ for every $S\subseteq N$, proving \ref{item:thermomaj} $\implies$
\ref{item:capacity2_charac}.
Conversely, since the target belongs to its own reachable region, $y(S)\leq C_\sigma(S)$, and hence \ref{item:capacity2_charac} $\implies$ \ref{item:capacity_charac}.

Clearly, \ref{item:capacity2_charac} $\implies$ \ref{item:capacity}.
For the reverse implication to thermomajorization, let $t_k=\gamma(S_k^{\pi^y})$.
Since $C_\sigma(S_k^{\pi^y}) =y(S_k^{\pi^y})=\Uyg(t_k)$, condition \ref{item:capacity} gives $\Uxg(t_k)\geq\Uyg(t_k)$ for $k=1,\ldots,n-1$. 
Both curves equal at $t=0$ and $t=1$.
Between consecutive target nodes, $\Uxg$ is concave and $\Uyg$ is affine, so the inequalities extend to all $t\in[0,1]$.
Thus \ref{item:capacity} $\implies$ \ref{item:thermomaj}.

Finally, Eq.~\eqref{eq:binary-capacity-response} gives
\begin{equation}
\DTO(\rho;K_k)-\DTO(\sigma;K_k)=\egap \bigl[C_\rho(S_k^{\pi^y}) -C_\sigma(S_k^{\pi^y})\bigr].
\label{eq:em-probe-difference}
\end{equation}
Since $\egap>0$, \ref{item:capacity} $\iff$ \ref{item:energy-storage}.
\end{proof}

\subsection{Constructive formula for the least element and universal first-contact smoothing}

\begin{proof}[Proof of Theorem~\ref{thm:construction}]
Write $g=\gxb=C_\rho+b$. Since $b$ is modular, $g$ inherits normalization and submodularity from $C_\rho$ (Theorem~\ref{thm:lower-base}), and
$B(g)=\Lgx+\vecb$ by Eq.~\eqref{eq:translated-base}. 
We construct a candidate $\vecz^\star\in B(g)$, show that it saturates the principal sets, and then identify it with $\xstar+\vecb$.

\textit{Supporting lines.}
By definition of the weighted principal sequence, the points
$(d(S_j),g(S_j))$ are the consecutive vertices of the greatest convex minorant
of the roof~\eqref{eq:roof-set}, and $\lambda_j$ is the slope of its $j$th
segment. A convex function lies above the line extending any of its linear
pieces, and the minorant lies below every roof point. Hence, for every
$T\subseteq N$ and $j=1,\ldots,m$,
\begin{equation}
g(T)\geq g(S_{j-1})+\lambda_j[d(T)-d(S_{j-1})].
\label{eq:em-roof-support}
\end{equation}
For $A\subseteq B_j$, taking $T=S_{j-1}\cup A$, so that
$d(T)-d(S_{j-1})=d(A)$, gives the marginal bound
\begin{equation}
\lambda_j\,d(A)\leq g(S_{j-1}\cup A)-g(S_{j-1}).
\label{eq:em-marginal}
\end{equation}

\textit{Candidate and feasibility.}
Define $z_i^\star=d_i\lambda_j$ for $i\in B_j$, so that $z_i^\star/d_i$ is
constant on each block and $z^\star(A)=\lambda_j d(A)$ for $A\subseteq B_j$.
For arbitrary $T\subseteq N$, splitting $T$ into the pieces $T\cap B_j$ and
applying Eq.~\eqref{eq:em-marginal} to each piece gives
\begin{equation}
\begin{aligned}
z^\star(T)&\leq\sum_{j=1}^m[g(S_{j-1}\cup(T\cap B_j))-g(S_{j-1})]\\
&\leq\sum_{j=1}^m[g(T\cap S_j)-g(T\cap S_{j-1})]=g(T).
\end{aligned}
\label{eq:em-feasible}
\end{equation}
The second inequality is submodularity in its diminishing-returns form:
adding $T\cap B_j$, which is disjoint from $S_{j-1}$, to the larger set
$S_{j-1}$ yields at most the increment obtained by adding it to
$T\cap S_{j-1}\subseteq S_{j-1}$. The final sum telescopes to
$g(T)-g(\emptyset)=g(T)$.
On the principal sets these bounds are tight: by Eq.~\eqref{eq:slopes},
$\lambda_\ell d(B_\ell)=g(S_\ell)-g(S_{\ell-1})$, so
\begin{equation}
z^\star(S_j)=\sum_{\ell=1}^j\lambda_\ell\,d(B_\ell)=g(S_j),
\qquad j=0,\ldots,m.
\label{eq:em-saturated}
\end{equation}
In particular, $z^\star(N)=g(N)$. Together with Eq.~\eqref{eq:em-feasible},
this gives $\vecz^\star\in B(g)$.

\textit{Identification.}
It suffices to compare $\vecz^\star$ with $\xstar+\vecb$ for a single strictly
convex cost. Let $Q(\vecz)=\frac12\sum_i z_i^2/d_i$, which is strictly convex
since $\vecd>0$. For $\vecz\in B(g)$, set $\delta_j=z(S_j)-z^\star(S_j)$.
Then $\delta_j\leq0$, because $z(S_j)\leq g(S_j)=z^\star(S_j)$, and
$\delta_0=\delta_m=0$. By convexity, $Q$ lies above its tangent plane at
$\vecz^\star$, whose gradient has components $z_i^\star/d_i=\lambda_j$ on
$B_j$. Therefore,
\begin{equation}
\begin{aligned}
Q(\vecz)-Q(\vecz^\star)
&\geq\sum_{j=1}^m\lambda_j[z(B_j)-z^\star(B_j)]\\
&=\sum_{j=1}^{m-1}(\lambda_j-\lambda_{j+1})\,\delta_j\geq0,
\end{aligned}
\label{eq:em-quadratic-optimality}
\end{equation}
where the equality uses $z(B_j)-z^\star(B_j)=\delta_j-\delta_{j-1}$ and
summation by parts, and the last inequality uses $\lambda_j<\lambda_{j+1}$ and
$\delta_j\leq0$. Hence $\vecz^\star$ is the unique minimizer of $Q$ on $B(g)$.
On the other hand, for $\phi(t)=t^2/2$, Eq.~\eqref{eq:general-shift-cost}
gives $F_{\vecd,\vecb,\phi}(\vecxp)=Q(\vecxp+\vecb)$, so by
Corollary~\ref{cor:universal_least} the vector $\xstar+\vecb$ also minimizes
$Q$ on $\Lgx+\vecb=B(g)$. By uniqueness, $\xstar=\vecz^\star-\vecb$, which is
Eq.~\eqref{eq:construction-main}.

\textit{Saturation.}
Finally, Eq.~\eqref{eq:em-saturated} gives
$x^\star_{\vecd,\vecb}(S_j)=z^\star(S_j)-b(S_j)=g(S_j)-b(S_j)=C_\rho(S_j)$,
which is Eq.~\eqref{eq:principal-capacity-saturation}.
\end{proof}

\begin{proof}[Proof of Corollary~\ref{cor:smoothing-equivalence}]
Fix $\vecd\in\Dno$ and a faithful error $\Edphi$, and let $\epsstar=\epsstar_{\vecd,\phi}(\rho\TO\sigma)$.
By Eq.~\eqref{eq:optimal-error}, the universal approximant $\xstary$ attains this error, so $\xstary\in\Lgx\cap \B_{\epsstar}^{\vecd,\phi}(\vecy)$.
For any $0\leq\varepsilon<\epsstar$, an intersection would imply the existence of a reachable state with error strictly below the minimum, a contradiction.
Since $\xstary$ is independent of $\phi$ at fixed $\vecd$, it provides a common first-contact state for all errors in the family.

By faithfulness, $\epsstar=0$ if and only if $\vecy\in\Lgx$, in which case the contact state is uniquely $\vecy$.
More generally, if $\phi$ is strictly convex, then $\Edphi(\cdot\,|\,\vecy)$ is strictly convex on the convex reachable region, and the contact state is unique. 
\end{proof}

\subsection{Trace-distance error from capacity and work-extraction deficits}

\begin{proof}[Proof of Theorem~\ref{thm:trace-distance-work-extraction}]
Write $D=\Dtr^\star(\rho\TO\sigma)$ and $P_k=S_k^{\pi^y}$.
For $\vec{w}\in\RR^n$ with $w(N)=0$, the positive and negative parts of
$\vec{w}$ have equal total weight, and $w(S)\le\sum_{i\in S}[w_i]_+$, where $[a]_+=\max\{0,a\}$. 
Hence
\begin{equation}
\tfrac12\norm{\vec{w}}_1=\textstyle\sum_{i}[w_i]_+\geq w(S)
\quad\forall S\subseteq N.
\label{eq:l1-positive-part}
\end{equation}

\textit{Proof of Eq.~\eqref{eq:trace-capacity-target-demand}.}
For every $\vecxp\in\Lgx$ and $S\subseteq N$,
\begin{equation}
\tfrac12\norm{\vecxp-\vecy}_1\geq y(S)-x'(S)\geq y(S)-C_\rho(S),
\end{equation}
where the first inequality is Eq.~\eqref{eq:l1-positive-part} with
$\vec{w}=\vecy-\vecxp$, and the
second follows from Eq.~\eqref{eq:thermal-population-capacity}.
Minimizing over $\vecxp$ and maximizing over $S$ gives
$D\geq\max_{S}[y(S)-C_\rho(S)]$.

Conversely, fix $\vecd\in\Dno$, let $\vecv=\xstary$, and
$S_-=\{i:v_i<y_i\}$. Then
\begin{equation}
D=\tfrac12\norm{\vecv-\vecy}_1=y(S_-)-v(S_-).
\end{equation}
The first equality follows from Eqs.~\eqref{eq:optimal-error} and \eqref{eq:optimal-trace-error}, since $E_{\vecd,\phi_1}(\vecv\,|\,\vecy)=\norm{\vecv-\vecy}_1$; the second is Eq.~\eqref{eq:l1-positive-part} with $\vec{w}=\vecy-\vecv$, since $[y_i-v_i]_+>0$ exactly for $i\in S_-$.

It remains to show that $\vecv$ saturates the capacity of $S_-$.
By Eq.~\eqref{eq:construction-main} with $\vecb=-\vecy$,
$v_i-y_i=d_i\lambda_j$ on $B_j$. Since $d_i>0$ and the slopes $\lambda_j$ strictly increase, $S_-$ is the union of the blocks with $\lambda_j<0$, that is, a cumulative principal set (possibly empty). Hence
Eq.~\eqref{eq:principal-capacity-saturation} gives $v(S_-)=C_\rho(S_-)$, and
\begin{equation}
D=y(S_-)-C_\rho(S_-)\leq\max_{S\subseteq N}[y(S)-C_\rho(S)].
\end{equation}
Both bounds prove Eq.~\eqref{eq:trace-capacity-target-demand}.

\textit{Proof of Eqs.~\eqref{eq:trace-capacity-deficit} and \eqref{eq:trace-target-capacity-gap}.}
Let $F(t)=\Uyg(t)-\Uxg(t)$, $t_k=\gamma(P_k)$, with $t_0=0$ and $t_n=1$, and
$M=\max_{0\le k\le n}F(t_k)$.
We use three facts:
(i) $C_\sigma(S)-C_\rho(S)=F(\gamma(S))$ for every $S\subseteq N$, by
Theorem~\ref{thm:lower-base} applied to $\rho$ and to $\sigma$;
(ii) $F\leq M$ on $[0,1]$, since on each $[t_{k-1},t_k]$ the curve $\Uyg$ is
affine and $\Uxg$ is concave, so $F$ is convex and attains its maximum at an
endpoint;
(iii) $y(S)\leq\Uyg(\gamma(S))=C_\sigma(S)$, with equality at every prefix
$P_k$, by Eq.~\eqref{eq:subset-below-curve} and Theorem~\ref{thm:lower-base}.
Hence, for every $S\subseteq N$,
\begin{equation}
y(S)-C_\rho(S)\leq C_\sigma(S)-C_\rho(S)\leq M,
\end{equation}
where we use (iii), (i), and (ii).
Maximizing over $S$ and using Eq.~\eqref{eq:trace-capacity-target-demand}, we obtain
$D\leq\max_{S\subseteq N}[C_\sigma(S)-C_\rho(S)]\leq M$.

On the other hand, for $k=0,\ldots,n$, 
\begin{equation}
y(P_k)-C_\rho(P_k)=C_\sigma(P_k)-C_\rho(P_k)=F(t_k),
\end{equation}
where we use (iii) and (i). Choosing $k$ with $F(t_k)=M$ and using
Eq.~\eqref{eq:trace-capacity-target-demand} gives $D\geq M$. Therefore
$D=\max_{S\subseteq N}[C_\sigma(S)-C_\rho(S)]=M$, proving
Eq.~\eqref{eq:trace-capacity-deficit}.
Since $\Uxg$ and $\Uyg$ both run from $(0,0)$ to $(1,1)$, $F(t_0)=F(t_n)=0$,
so dropping $k=0$ does not change the maximum: $M=\max_{1\le k\le n}F(t_k)$,
which by (i) is Eq.~\eqref{eq:trace-target-capacity-gap}.

\textit{Proof of Eq.~\eqref{eq:trace-from-work}.}
Eqs.~\eqref{eq:em-probe-difference} and \eqref{eq:trace-target-capacity-gap} give Eq.~\eqref{eq:trace-from-work},
since $\egap>0$.
\end{proof}


\section{Constructive algorithms and complexity analysis}
\label{app:algorithm}

We now describe how to compute the weighted principal sequence of Theorem~\ref{thm:construction}, and hence the universal least element, without enumerating the exponentially many subsets of energy levels.
The construction has two ingredients: supporting lines identify the relevant points of the capacity roof, and the piecewise-linear Lorenz curve makes each supporting-line search explicit. 
We first explain the geometry and the recursive construction, then derive its minimization oracle. 
A further simplification applies to Gibbs- and target-weighted approximation, for which a single ordering of the target populations suffices. 
The proofs of global correctness and the complexity bounds are given at the end.

Throughout, let $N=\{1,\ldots,n\}$,
$\vecx\in\Dn$, $\vecg,\vecd\in\Dno$, and $\vecb\in\RR^n$.
We abbreviate the shifted capacity function in
Eq.~\eqref{eq:gxb} as
\begin{equation}
 g(S)=C_\rho(S)+b(S)=\Uxg\bigl(\gamma(S)\bigr)+b(S).
 \label{eq:supp-g}
\end{equation}
Theorem~\ref{thm:lower-base} implies that $g$ is normalized ($g(\emptyset)=0$) and submodular, although the modular shift $b(S)$ need not preserve monotonicity.
The discrete capacity roof $\capacityroof$ comprises the points $(d(S),g(S))$, $S\subseteq N$.
Its greatest convex minorant (GCM) is the lower boundary of their convex hull, regarded as a convex piecewise-linear function on $[0,1]$. 
Its successive vertices determine blocks $B_j=S_j\setminus S_{j-1}$ and slopes $\lambda_j$, from which the universal optimizer follows directly as $x^\star_{\vecd,\vecb,i}=d_i\lambda_j-b_i$ for $i\in B_j$.
Thus the computational task is to find these blocks and slopes.

\subsection{Supporting lines and recursive construction}

To identify the supporting lines of the capacity roof, consider a fixed slope $\lambda\in\mathbb R$ and define
\begin{equation}
 Q_\lambda(S)=g(S)-\lambda d(S).
 \label{eq:supp-parametric}
\end{equation}
Geometrically, $Q_\lambda(S)$ is the vertical intercept of the line of slope $\lambda$ passing through the roof point $(d(S),g(S))$.

For this line to lie below every roof point, its intercept must not exceed $Q_\lambda(S)$ for any
$S\subseteq N$. 
Thus, the highest possible intercept is
\begin{equation}
 q(\lambda)=\min_{S\subseteq N}Q_\lambda(S).
 \label{eq:supp-support-intercept}
\end{equation}
Since the roof contains finitely many points, this minimum is attained. 
Consequently, the line $\ell_\lambda(t)=\lambda t+q(\lambda)$ lies below the entire roof and touches it precisely at the points associated with minimizers of $Q_\lambda$.

As $\lambda$ varies, these lines recover the greatest convex minorant: every lower supporting line lies below it, while each of its linear segments belongs to this family.
Rather than searching over all slopes, our algorithm identifies the finitely many segment slopes by tracking the corresponding minimizing sets.

The key to this construction is the submodularity of $Q_\lambda$. 
Indeed, if $S$ and $T$ both attain the minimum, submodularity gives 
\begin{equation}
Q_\lambda(S\cap T)+Q_\lambda(S\cup T)\leq Q_\lambda(S)+Q_\lambda(T),    
\end{equation}
so both $S\cap T$ and $S\cup T$ are minimizers as well. 
Hence, there exist a unique largest and a unique smallest minimizer by inclusion, denoted $\splus(\lambda)$ and $\sminus(\lambda)$, respectively.
The same property holds when minimization is restricted to an interval $A\subseteq S\subseteq B$.

More importantly, the minimizers are nested as the slope increases: for $\lambda<\mu$, every minimizer
of $Q_\lambda$ is contained in every minimizer of $Q_\mu$, as established by the following lemma.
This nesting property allows the relevant contact sets to be organized into a chain, from which the segments of the convex minorant can be recovered.

\begin{lemma}[Nesting of parametric minimizers]
\label{lem:supp-nesting}
If $\lambda<\mu$, every minimizer of $Q_\lambda$ is contained in
every minimizer of $Q_\mu$. In particular,
$\splus(\lambda)\subseteq\sminus(\mu)\subseteq\splus(\mu)$.
\end{lemma}
\begin{proof}
Let $S$ minimize $Q_\lambda$ and $T$ minimize $Q_\mu$, and set
$I=S\cap T$, $J=S\cup T$. Submodularity of $g$ and modularity of
$d$ imply
\begin{equation}
 Q_\lambda(S)+Q_\mu(T)\ge
 Q_\lambda(I)+Q_\mu(J)
 +(\mu-\lambda)d(S\setminus T).
 \label{eq:supp-uncrossing}
\end{equation}
By optimality, the left-hand side is no larger than the first two
terms on the right. Hence
$(\mu-\lambda)d(S\setminus T)\le0$.
Since $\mu>\lambda$ and $d_i>0$ for all $i$, we have $S\subseteq T$.
\end{proof}

We use this nesting to recover the GCM recursively. 
For two nested sets $A\subsetneq B$, the slope of the chord joining their roof points is
\begin{equation}
 \lambda(A,B)=\frac{g(B)-g(A)}{d(B)-d(A)}.
 \label{eq:supp-chord}
\end{equation}
The denominator is positive because $\vecd>0$. For any $\mu$,
\begin{equation}
 Q_\mu(B)-Q_\mu(A) =[d(B)-d(A)][\lambda(A,B)-\mu].
 \label{eq:supp-chord-comparison}
\end{equation}
At $\mu=\lambda(A,B)$ the endpoints tie. We therefore minimize $Q_{\lambda(A,B)}$ over the interval between them,
\begin{equation}
 \min_{A\subseteq S\subseteq B}Q_{\lambda(A,B)}(S),
 \label{eq:supp-constrained}
\end{equation}
and denote its largest minimizer by inclusion as $M(A,B;\lambda(A,B))$.
The outcome has a direct interpretation:
\begin{itemize}
 \item If $M=B$, no intermediate roof point in the interval lies  below the chord. 
 We accept $B\setminus A$ as one block, with slope $\lambda(A,B)$.
 \item If $M\ne B$, the tied endpoints cannot minimize the interval problem. Therefore
 \begin{equation}
 A\subsetneq M\subsetneq B,\quad  Q_{\lambda(A,B)}(M)<Q_{\lambda(A,B)}(A).
 \label{eq:supp-split}
 \end{equation}
 The point corresponding to $M$ lies strictly below the chord, so we recursively examine $(A,M)$ and $(M,B)$.
\end{itemize}
Choosing the largest minimizer is important when several points are collinear: a supporting chord is accepted rather than split into redundant segments. 
The restricted test is sufficient because, as proved in Lemma~\ref{lem:supp-recursion}, every interval encountered by the recursion contains the relevant global minimizers.

\begin{algorithm}[H]
\caption{$\CapRoof(A,B)$: recursive capacity-roof construction}
\label{alg:supp-roof}
\begin{algorithmic}[1]
\Require $A\subsetneq B\subseteq N$, with $g$ and $d$ fixed.
\Ensure Ordered block--slope pairs joining $A$ to $B$.
\State $\displaystyle\lambda\gets\frac{g(B)-g(A)}{d(B)-d(A)}$.
\State $M\gets\MaxMin(A,B;\lambda)$.
\If{$M=B$}
 \State \Return $[(B\setminus A,\lambda)]$.
\Else
 \State \Return $\CapRoof(A,M)$ concatenated with $\CapRoof(M,B)$.
\EndIf
\end{algorithmic}
\end{algorithm}

The initial call is $\CapRoof(\emptyset,N)$. 
Its output is the sequence of block--slope pairs $(B_j,\lambda_j)$. 
The next subsection shows how to implement the oracle $\MaxMin$ using only sorting, arithmetic, and a linear sweep through the Lorenz segments.

\subsection{Efficient minimization from the Lorenz curve}

The apparent difficulty in Eq.~\eqref{eq:supp-constrained} is the minimization over subsets. 
The special form of $g$ removes it, since the piecewise-linear structure of the Lorenz curve reduces this problem to elementary additive minimizations.

Since the relative Lorenz curve is concave and piecewise linear, each of its linear segments extends to a supporting line lying above the entire curve.
Moreover, at every point at least one such line coincides with the curve. 
Hence,
\begin{equation}
 \Uxg(t)=\min_{1\le r\le K}\{a_rt+c_r\},
 \qquad a_1>\cdots>a_K,\quad K\le n.
 \label{eq:supp-lorenz-affine}
\end{equation}
The slopes $a_r$ are the distinct values of $x_i/\gamma_i$ in decreasing order, with consecutive equal slopes merged. 
The intercept of each line is $c_r=\Uxg(u_r)-a_ru_r$, where $u_r$ is the left endpoint of the corresponding Lorenz segment.

For fixed $\lambda$, define
\begin{equation}
 \eta_i^{(r)}(\lambda)=a_r\gamma_i+b_i-\lambda d_i.
 \label{eq:supp-eta}
\end{equation}
Substitution into Eq.~\eqref{eq:supp-parametric} gives
\begin{equation}
 Q_\lambda(S)=\min_{1\le r\le K}
 \left\{c_r+\sum_{i\in S}\eta_i^{(r)}(\lambda)\right\}.
 \label{eq:supp-modular-reduction}
\end{equation}
For each fixed $r$, the expression in braces is modular: each level contributes independently. 
Over $A\subseteq S\subseteq B$, the indices in $A$ must be included, those outside $B$ are excluded, and each remaining index is included precisely when its coefficient is nonpositive. 
Thus the largest minimizer and the minimum of the $r$th modular function are
\begin{align}
 S_r&=A\cup\{i\in B\setminus A:\eta_i^{(r)}(\lambda)\le0\},
 \label{eq:supp-Sr}\\
 q_r&=c_r+\sum_{i\in A}\eta_i^{(r)}(\lambda)
 +\sum_{i\in B\setminus A}\min\{0,\eta_i^{(r)}(\lambda)\}.
 \label{eq:supp-qr}
\end{align}
The weak inequality includes zero-coefficient indices and is necessary for maximality.

It is useful to express this decision as a threshold test. Set
\begin{equation}
 t_i(\lambda)=\frac{\lambda d_i-b_i}{\gamma_i},
 \qquad i\in B\setminus A,
 \label{eq:supp-threshold}
\end{equation}
so that
$\eta_i^{(r)}(\lambda)=\gamma_i[a_r-t_i(\lambda)]$.
As $\gamma_i>0$, the inclusion rule becomes
$t_i(\lambda)\ge a_r$. Since the $a_r$ decrease with $r$,
$S_1\subseteq\cdots\subseteq S_K$. This nesting also specifies how
to resolve ties between different affine pieces.

\begin{lemma}[Explicit maximal-minimizer oracle]
\label{lem:supp-oracle}
Let $S_r$ and $q_r$ be given by
Eqs.~\eqref{eq:supp-Sr}--\eqref{eq:supp-qr}, and put
\begin{equation}
 q^\star=\min_{1\le r\le K}q_r,\qquad
 r^\star=\max\{r:q_r=q^\star\}.
 \label{eq:supp-last-tie-index}
\end{equation}
Then the largest minimizer of $Q_\lambda$ over
$A\subseteq S\subseteq B$ is
\begin{equation}
 M(A,B;\lambda)=S_{r^\star}.
 \label{eq:supp-last-tie}
\end{equation}
\end{lemma}
\begin{proof}
The finite minima over $S$ and $r$ in
Eq.~\eqref{eq:supp-modular-reduction} commute, so the constrained
minimum of $Q_\lambda$ is $q^\star$. At $S_{r^\star}$ we have
$Q_\lambda(S_{r^\star})\le q_{r^\star}=q^\star$, hence equality.
Conversely, let $S$ minimize $Q_\lambda$, and choose a piece $r$
active at $S$. Then
\begin{equation*}
 q^\star\le q_r\le c_r+\sum_{i\in S}\eta_i^{(r)}(\lambda)
 =Q_\lambda(S)=q^\star.
\end{equation*}
Thus $q_r=q^\star$ and $S$ minimizes the $r$th modular function.
It follows that $S\subseteq S_r\subseteq S_{r^\star}$, because
$r\le r^\star$ and the $S_r$ are nested. Hence $S_{r^\star}$
contains every minimizer.
\end{proof}

To evaluate all $q_r$ efficiently, sort the $p=|B\setminus A|$ indices as $t_{i_1}(\lambda)\ge\cdots\ge t_{i_p}(\lambda)$ and set $k_r=\#\{i\in B\setminus A:t_i(\lambda)\ge a_r\}$. 
Then
\begin{align}
 S_r&=A\cup\{i_1,\ldots,i_{k_r}\},
 \label{eq:supp-Sr-prefix}\\
 q_r&=c_r+a_r\left[\gamma(A)+
 \sum_{j=1}^{k_r}\gamma_{i_j}\right]\notag\\
 &\quad +b(A)-\lambda d(A)
 +\sum_{j=1}^{k_r}(b_{i_j}-\lambda d_{i_j}).
 \label{eq:supp-qr-prefix}
\end{align}
The values $k_r$ are nondecreasing, so each index enters at most once. 
Algorithm~\ref{alg:supp-oracle} carries out this sweep.
Its accumulators $G$ and $h$ store the Gibbs weight and the shifted linear contribution of the included prefix, while $k$ is its size.

\begin{algorithm}[H]
\caption{$\MaxMin(A,B;\lambda)$: sorted-threshold oracle}
\label{alg:supp-oracle}
\begin{algorithmic}[1]
\Require $A\subsetneq B$, $\lambda$, and $(a_r,c_r)_{r=1}^K$ with $a_1>\cdots>a_K$.
\Ensure Largest minimizer of $Q_\lambda$ on $A\subseteq S\subseteq B$.
\State $G_A\gets\gamma(A)$; $h_A\gets b(A)-\lambda d(A)$.
\State Sort $B\setminus A=\{i_1,\ldots,i_p\}$ by $t_{i_1}(\lambda)\ge\cdots\ge t_{i_p}(\lambda)$.
\State $k\gets0$; $G\gets0$; $h\gets0$; $q^\star\gets+\infty$; $k^\star\gets0$.
\For{$r=1,\ldots,K$}
 \While{$k<p$ \textbf{and} $t_{i_{k+1}}(\lambda)\ge a_r$}
  \State $k\gets k+1$; $G\gets G+\gamma_{i_k}$; $h\gets h+b_{i_k}-\lambda d_{i_k}$.
 \EndWhile
 \State $q\gets c_r+a_r(G_A+G)+h_A+h$.
 \If{$q\le q^\star$}
  \State $q^\star\gets q$; $k^\star\gets k$.
 \EndIf
\EndFor
\State \Return $A\cup\{i_1,\ldots,i_{k^\star}\}$.
\end{algorithmic}
\end{algorithm}

At iteration $r$, the accumulated values evaluate $q_r$ in Eq.~\eqref{eq:supp-qr-prefix}. 
The two weak comparisons serve different purposes: $t_i\ge a_r$ includes zero coefficients, and $q\le q^\star$ retains the last minimizing affine piece. 
Together they ensure that the output is exactly the maximal minimizer.

\subsection{Fixed-order prefix construction for two natural geometries}

The general oracle sorts thresholds that may depend on $\lambda$. 
For the two physically relevant choices of comparison weights, Gibbs ($\vecd=\vecg$) and target ($\vecd=\vecy$), the relevant ordering is fixed across
all slopes. The entire GCM can then be obtained from only $n+1$ roof points by an ordinary lower convex-hull construction.

\begin{corollary}[Fixed-order prefix construction]
\label{cor:supp-prefix}
Let $\vecb=-\vecy$, and consider the target $\vecg$-ordering $\pi^y$ and its prefix sets $S_k^{\pi^y}$, defined in the main text.
If $\vecd=\vecg$, or if $\vecd=\vecy\in\Dno$, the GCM of $\capacityroof$ equals the GCM of the $n+1$ prefix points
\begin{equation}
 \left( d(S_k^{\pi^y}),C_\rho(S_k^{\pi^y})-y(S_k^{\pi^y}) \right), \quad k=0,\ldots,n.
 \label{eq:supp-prefix-points}
\end{equation}
In either case, the universal optimizer is computable in $\BigO(n\log n)$ arithmetic and comparison operations.

More generally, when $\vecd=\vecg$ and $\vecb$ is arbitrary, the same conclusion holds with prefixes ordered by nonincreasing values of $-b_i/\gamma_i$ and ordinates $g(S_k^{\pi^y})$.
\end{corollary}

\begin{proof}
By Eqs.~\eqref{eq:supp-threshold} and \eqref{eq:supp-last-tie}, the maximal minimizer of $Q_\lambda$ is a threshold prefix in the ordering of
\begin{equation}
 t_i(\lambda)=\frac{\lambda d_i-b_i}{\gamma_i}.   
\end{equation}
 
For $\vecb=-\vecy$ and $\vecd=\vecg$, these thresholds become 
\begin{equation}
 t_i(\lambda)=\lambda+\frac{y_i}{\gamma_i},   
\end{equation}
so their ordering is $\pi^y$, independently of $\lambda$.

For $\vecb=-\vecy$ and $\vecd=\vecy\in\Dno$, the thresholds are instead
\begin{equation}
 t_i(\lambda)=(1+\lambda)\frac{y_i}{\gamma_i}.   
\end{equation}
 
If $\lambda>-1$, their ordering is again $\pi^y$.
If $\lambda\le-1$, then
\begin{equation}
 Q_\lambda(S)=C_\rho(S)-(1+\lambda)y(S)\ge0,   
\end{equation}
 with strict inequality for nonempty $S$, because $\gamma_i>0$ implies $C_\rho(S)>0$ whenever $S\ne\emptyset$. 
Thus $\emptyset$ is the unique minimizer.

In both geometries, for every $\lambda$ there exists a target prefix $S_k^{\pi^y}$ that minimizes $Q_\lambda$ globally. 
Hence
\begin{equation}
\min_{S\subseteq N}  [g(S)-\lambda d(S)] =  \min_{0\le k\le n}  \bigl[ g(S_k^{\pi^y}) -\lambda d(S_k^{\pi^y}) \bigr]
\label{eq:supp-prefix-support}
\end{equation}
for every real $\lambda$. 
The two point collections therefore have the same supporting lines from below and the same GCM.

For the generalization to arbitrary $\vecb$ with $\vecd=\vecg$, the thresholds become $t_i(\lambda)=\lambda-b_i/\gamma_i$.
Their ordering is likewise independent of $\lambda$, and the same argument applies to the corresponding prefix sets.

Sorting the relevant ratios costs $\BigO(n\log n)$; the prefix points can then be evaluated in $\BigO(n)$ after construction of the Lorenz curve. 
Their lower convex hull is obtained by a linear stack sweep, retaining only vertices where the slope increases.
The final block--slope pairs give the optimizer through Eq.~\eqref{eq:construction-main}.
\end{proof}

For clarity, the prefix method requires no submodular-minimization oracle. 
After computing the points in Eq.~\eqref{eq:supp-prefix-points}, process them in increasing order of $k$ and maintain a stack
of selected prefix indices. 
Whenever the last three indices $a<b<c$ satisfy
\begin{equation}
\lambda(S_a^{\pi^y},S_b^{\pi^y})\geq \lambda(S_b^{\pi^y},S_c^{\pi^y}),
\label{eq:supp-prefix-stack}
\end{equation}
remove the middle index $b$, where the chord slope $\lambda(A,B)$ is defined in Eq.~\eqref{eq:supp-chord}.
Repeating this test leaves the successive GCM vertices; equality removes collinear intermediate
points. In particular, the selected principal sets need not be the same for Gibbs and target weights,
even though both are drawn from the same target-prefix chain.

This fixed-prefix property should not be inferred for general $\vecd$. For $\vecb=-\vecy$, the
thresholds are
\begin{equation}
 t_i(\lambda)=\frac{y_i+\lambda d_i}{\gamma_i},   
\end{equation}
 and their ordering may change with the supporting slope. 
 Although the minimum of $g(S)=C_\rho(S)-y(S)$ always occurs at a target prefix, the first slope of the GCM is instead
\begin{equation}
 \lambda_1=  \min_{S\ne\emptyset}\frac{g(S)}{d(S)},
 \label{eq:supp-first-slope}
\end{equation}
which need not be attained by a prefix of the target ordering. 
This is why the general recursive construction remains necessary.

\subsection{Correctness of the recursive construction}

The remaining issue is globality: the recursion in
Algorithm~\ref{alg:supp-roof} minimizes over restricted intervals $A\subseteq S\subseteq B$, whereas the GCM is defined by all subsets of $N$. We show that every interval encountered is bounded by two globally optimal parametric sets, so the restriction never discards a relevant lower roof point.

\begin{lemma}[Global correctness of the chord recursion]
\label{lem:supp-recursion}
The initial call to Algorithm~\ref{alg:supp-roof} returns block--slope
pairs $(B_j,\lambda_j)_{j=1}^m$ whose cumulative unions satisfy
\begin{equation}
 \emptyset=S_0\subsetneq S_1\subsetneq\cdots\subsetneq S_m=N,
 \qquad m\le n,
 \label{eq:supp-chain}
\end{equation}
with strictly increasing slopes
\begin{equation}
 \lambda_j=\frac{g(S_j)-g(S_{j-1})}{d(S_j)-d(S_{j-1})},
 \qquad\lambda_1<\cdots<\lambda_m.
 \label{eq:supp-slopes}
\end{equation}
Moreover, $S_{j-1}=\sminus(\lambda_j)$ and
$S_j=\splus(\lambda_j)$, so the corresponding chords support the
\emph{full} roof. The points $(d(S_j),g(S_j))$ are the successive
vertices of its GCM, with collinear intermediate points suppressed.
\end{lemma}
\begin{proof}
\emph{Invariant and initialization.} Define $H(A,B)$ to mean that
$A\subsetneq B$ and that there exist $\lambda_A<\lambda_B$ with
$A=\splus(\lambda_A)$ and $B=\splus(\lambda_B)$.
The initial pair satisfies $H(\emptyset,N)$: since all $d_i>0$,
$Q_\mu$ is uniquely minimized by $\emptyset$ for sufficiently
negative $\mu$ and by $N$ for sufficiently positive $\mu$.

\emph{Why the interval oracle is global.} Suppose $H(A,B)$ and put
$\lambda=\lambda(A,B)$. Since $A=\splus(\lambda_A)$ and $B\ne A$,
we have $Q_{\lambda_A}(B)>Q_{\lambda_A}(A)$, whereas
$Q_{\lambda_B}(B)\le Q_{\lambda_B}(A)$. By
Eq.~\eqref{eq:supp-chord-comparison}, this yields
$\lambda_A<\lambda\le\lambda_B$. Lemma~\ref{lem:supp-nesting}
then shows that every global minimizer of $Q_\lambda$ contains
$A$ and that its largest minimizer lies in $B$. Thus a global
minimizer is available inside the constrained interval, and
\begin{equation}
 M(A,B;\lambda)=\splus(\lambda).
 \label{eq:supp-global-oracle}
\end{equation}
This identity is the reason why restricted recursive searches
recover the global GCM.

\emph{Terminal chords and preservation of the invariant.}
If $M=B$, both endpoints are global minimizers of $Q_\lambda$.
Moreover, every minimizer at $\lambda$ contains $A$ by nesting,
so $A=\sminus(\lambda)$; likewise $B=\splus(\lambda)$.
For every $S\subseteq N$, then,
$g(S)\ge g(A)+\lambda[d(S)-d(A)]$: the terminal chord is a
global supporting line.
If instead $M\ne B$, Eq.~\eqref{eq:supp-split} gives
$A\subsetneq M\subsetneq B$. We also have $\lambda<\lambda_B$,
for otherwise Eq.~\eqref{eq:supp-global-oracle} would imply
$M=\splus(\lambda_B)=B$. Since $M=\splus(\lambda)$, the two
children satisfy $H(A,M)$ and $H(M,B)$, with parameter intervals
$(\lambda_A,\lambda)$ and $(\lambda,\lambda_B)$, respectively.

\emph{Induction and recovery of the GCM.}
Each split strictly decreases $|B\setminus A|$ in both children,
so induction proves termination. Furthermore, when the parent chord
of slope $\lambda$ is split at $M$, Eq.~\eqref{eq:supp-split}
implies
$\lambda(A,M)<\lambda<\lambda(M,B)$.
Together with the invariant on the two children, this yields
terminal chords with strictly increasing slopes and endpoints
$\sminus(\lambda_j)$ and $\splus(\lambda_j)$.
For the initial call this gives
Eqs.~\eqref{eq:supp-chain}--\eqref{eq:supp-slopes}.
Because every block is nonempty, $m\le n$.
Let $h$ interpolate linearly between the returned roof points.
Its slopes increase strictly, so $h$ is convex. Each segment is
globally supporting; hence $h$ is below every roof point.
Conversely, any convex minorant $v$ of the roof satisfies
$v(d(S_j))\le g(S_j)=h(d(S_j))$ at every returned vertex.
Convexity places $v$ below the chord joining each consecutive
pair of such vertices, so $v\le h$ everywhere. Therefore $h$ is
the greatest convex minorant.

\emph{No omitted parametric transitions.}
For $\lambda_j<\lambda<\lambda_{j+1}$, nesting yields
$S_j=\splus(\lambda_j)\subseteq\splus(\lambda)
\subseteq\sminus(\lambda_{j+1})=S_j$.
Outside the first and last transition parameters, the extremal
minimizers are $\emptyset$ and $N$. Thus the maximal minimizers
visit exactly the returned cumulative sets, proving that this is
the global weighted principal sequence.
\end{proof}

The initial call therefore returns the blocks and slopes of Theorem~\ref{thm:construction}, and the universal least element is obtained componentwise as
\begin{equation}
 x^\star_{\vecd,\vecb,i}=d_i\lambda_j-b_i,  \quad i\in B_j=S_j\setminus S_{j-1}.
 \label{eq:supp-output}
\end{equation}
Its feasibility, saturation of the cumulative capacity constraints, and universal optimality follow from
Theorem~\ref{thm:construction} and Corollary~\ref{cor:universal_least}.

\subsection{Computational complexity}

The affine pieces of $\Uxg$ are computed by sorting the ratios $x_i/\gamma_i$ and accumulating their Lorenz nodes, in $\BigO(n\log n)$ arithmetic and comparison operations.
For one call to Algorithm~\ref{alg:supp-oracle}, let $p=|B\setminus A|$ and recall that $K\le n$. 
Computing the thresholds and the sums over $A$ takes $\BigO(n)$ operations, sorting takes $\BigO(p\log p)$, and the sweep takes
$\BigO(p+K)$. 
The oracle therefore costs
\begin{equation}
 \BigO(n+p\log p+K)=\BigO(n\log n).
 \label{eq:supp-oracle-cost}
\end{equation}
Computing the chord slope adds at most $\BigO(n+K)$ operations, so every recursive call costs $\BigO(n\log n)$.

If the resulting chain has $m$ nonempty blocks, the recursion tree is full binary with $m$ terminal nodes and $m-1$ internal nodes.
Since $m\le n$, there are at most $2n-1$ calls. 
Including Lorenz preprocessing gives
\begin{equation}
 T_{\mathrm{general}}(n)=\BigO(n^2\log n).
 \label{eq:supp-complexity}
\end{equation}
No enumeration of $2^n$ roof points or general-purpose submodular minimization is required.

For the fixed-order cases of Corollary~\ref{cor:supp-prefix}, the ratio sorting and Lorenz preprocessing together take $\BigO(n\log n)$. 
The prefix weights are cumulative, and their
Lorenz ordinates are evaluated in $\BigO(n)$ total time by a single sweep through the Lorenz segments as $\gamma(P_k)$ increases.
The standard stack construction of the lower convex hull likewise takes $\BigO(n)$ time. 
Hence
\begin{equation}
 T_{\mathrm{prefix}}(n)=\BigO(n\log n),
 \label{eq:supp-prefix-complexity}
\end{equation}
for Gibbs-weighted approximation and, for a full-support target, target-weighted approximation. 
This improvement exploits the fixed ordering of the relevant prefixes, although it does not extend to arbitrary comparison weights.


\bibliography{references}

\end{document}